\documentclass[11pt,twoside]{article}
\usepackage{asp2006}
\usepackage{psfig}
\usepackage{epsf}
\usepackage{graphics}
\usepackage{lscape}
\def\edcomment#1{\iffalse\marginpar{\raggedright\sl#1\/}\else\relax\fi}
\reversemarginpar

\begin{document}
\title{The formation of extreme mass ratio binary stars: a tribute to Eduardo Delgado Donate}
\author{C. J. Clarke}
\affil{Instite of Astronomy, Cambridge, CB3 OHA, U.K.}

\begin{abstract}
Eduardo Delgado was due to have presented a poster at this meeting
on his latest results on the formation of extreme mass ratio binaries.
Tragically, Eduardo was among those killed in a hiking  accident
in Tenerife earlier this year. As  his PhD supervisor, 
and as a longstanding collaborator, the organisers of
this  meeting kindly invited me to incorporate a report on his most recent work
into a more general tribute to his life and work.
  I will reflect on Eduardo's scientific career, the problems that
motivated him and  his achievements,   focusing  particularly on a  problem
which had intrigued  us both for several years and on which Eduardo
was making important progress at the time of his death. Finally, I will
mention the personal qualities that Eduardo brought to his work and
the acute sense of loss that is shared by all those - friends and collaborators
- who were  privileged to  know him.

\end{abstract}

\vspace{-0.5cm}
\section{Cambridge days: predictions and puzzles in multiple star formation}
Eduardo came to Cambridge from his native Tenerife in 2000, in order to work
on hydrodynamic (SPH) star formation simulations under my supervision. He
was  funded by an E.U. 
PhD studentship as part of the `Young Stellar Cluster' Network. Anyone  who has
been
involved in these Networks will know that   it is not always
easy to find students of suitable calibre to fill these posts, given
their restrictive nationality and residence requirements, and can therefore
imagine my delight at receiving such a strong application
from Eduardo.
He went on to fulfil my high
expectations, quickly developing into a fine numericist and astrophysicist.
(I should add that I was often struck by the fact  that Eduardo's written
English was far more eloquent than anything that his British contemporaries
could manage - although I'm afraid that this says something not only
about Eduardo but also about the British educational system!)

\begin{figure}[!ht]
\plotone{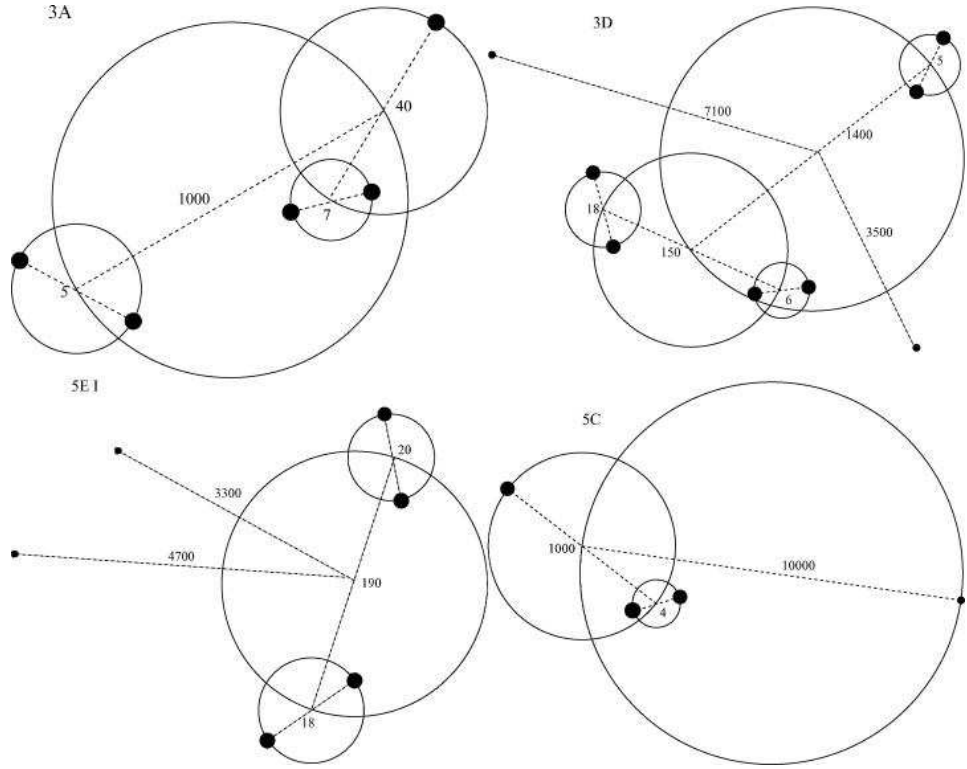}
\caption{ Examples of some of the system architectures of complex
multiple star systems formed in the hydrodynamical simulations of
Delgado et al 2004. The numbers refer to separations in A.U..}
\end{figure}

  Eduardo arrived in Cambridge when Matthew Bate was still a postdoc
there and Matthew was a great help to Eduardo in helping him to master
his SPH code. This was an exciting time for Eduardo to be working in
numerical star formation,  as this was the time that Matthew was producing
the first turbulent fragmentation calculations which have generated so
much interest (and controversy!) ever since.  It was clearly going
to be a challenge for Eduardo to do anything in this field which
was going to have a distinct impact. I think the collection of well cited
papers that Eduardo produced during his thesis shows that he met this
challenge most successfully.

 Eduardo became known in the field for his series of turbulent fragmentation
calculations which explored the production of multiple star systems
and brown dwarfs in small N ensembles. 
My own  vote for the part of his thesis that will be his  biggest contribution 
to the field
would go to his calculations which were  the first to  make 
predictions about the properties of higher order multiple
systems (see some of the complex system architectures generated in
Figure 1).  
He was able to do this because rather than
focusing his computational effort on system scale (as Matthew Bate
did with his
$50$ solar mass calculation), Eduardo followed small ($5$ solar mass) clouds
and - in compensation - was able to follow the calculation for much longer.
Thus  Eduardo's simulations could follow
even wide  
systems to the point of dynamical stability. Evidently this is important, not
only because wide binaries are numerous in nature (half of all binaries with
solar type primaries are wider than $30$ A.U.) but because in this way one
can start to make pronouncements about higher order multiples (which 
necessarily involve wide, as well as close, components). 

 Thus  Eduardo entered  uncharted waters in terms of simulations (and
indeed his paper on this, Delgado et al  2004 is still
the only one on this topic in existence). Moreover, the characterisation of
multiple star systems is far from complete observationally, as it is a
laborious 
task to collate complete statistics on components with a wide
dynamic range of separations (e.g. Tokovinin et al 2006).
Indeed, the motivation for  such observational studies has
always been undermined by the fact that - 
until Eduardo's
work - there were no theoretical  predictions with which to compare
the data.

  This brings me to one
of Eduardo's most important characteristics as a scientist, that in addition
to being
technically accomplished as a numericist, 
he had a strong interest in - and respect for - observations. One would
never hear Eduardo implying that observations much be wrong if they
contradicted his simulations - or that observers should simply go and
`look harder'. He understood, better than most, that it's actually
the areas of disagreement that illuminate your calculations. And, moreover,
he understood that the range of diagnostic information that is contained
in multiple star statistics would ultimately provide a very stringent -
arguably the most stringent - test of star formation theories.

  Thus Eduardo became  intrigued by an obvious area of discrepancy
between observations and all star formation simulations.
In the simulations,  all the binaries
were within a factor of two of being
of unit mass ratio (with the exception of a few weakly bound very low
mass `outliers' at large separations). Moreover, this result extended to
every level of the hierarchy (i.e. in a triple, a nearly equal mass binary
would be bound to a third star with mass nearly equal to the binary).
Evidently  nature does not work like that! It is true that there are categories
of binaries that favour more nearly equal masses, arguably  OB stars (Garcia \& Mermilliod
2001), very low mass binaries (Bouy et al 2003) and short period spectroscopic
binaries among solar type stars (Mazeh et al 1992).  
However, for the bulk of solar type binaries,
the median mass ratio  is
$\sim 0.4$  (Duquennoy \& Mayor 1991), a result  which - 
given that it is based
partly on pairs 
from  the visual binary literature  which are almost
certainly selectively incomplete at low
mass ratios  - must, if anything, {\it underestimate} the incidence
of pairs with very disparate masses.

  Early in 2003, Eduardo  started writing a paper with me, alas never
to be published, entitled `On the problem of forming extreme mass ratio
binaries,. I told him to start by writing a few pages setting up the
problem (which he did) and then, by the time he'd written that, his simulations
would have run to the point where he could unveil the solution (or at
any rate {\it a} solution)
in the concluding sections. But the latter proved to be impossible. None of
the fixes to initial conditions which he attempted  (such as adjusting the
power spectrum of the initial velocity field or changing the geometry)
were ultimately successful. 
He would excitedly tell me that yes, he'd managed to make a binary
with q  of $0.1$, only to return a week later to tell me that, no,
it had now evolved to q of $ 0.7$.... Thus it turned out 
that the problem was not  the
initial creation of such systems, but the effect of continued accretion
onto the protobinary, which drove up the mass ratio. 

It has been known since at least the work of Artymowicz
1981 (and note that Eduardo went on to spend two years in Stockholm
working with Pawel Artymowicz as a postdoc, although 
never on this problem) that accretion
of gas whose specific angular momentum exceeds that of the binary leads to
preferential accretion onto the secondary. This commonsense result, 
which was apparent in these early  ballistic calculations and 
was backed up by the 
subsequent
SPH simulations  of Bate \& Bonnell 1997 and Bate 2000,
occurs simply because the secondary's Roche lobe is further
from the system's centre of mass, and is thus more likely to intercept high
angular momentum material. Since in the case of any plausible binary formation
scenario, higher angular momentum material will fall in later, 
the binary mass ratio, q, should therefore rise with time.
 
\section{Numerical controversy: latest results}

While Eduardo and I abandonned this question, discouraged, (and Eduardo
took up a position in Stockholm where he  worked  mainly on codes for planet formation), there
meanwhile appeared the paper of Ochi, Sugimoto \& Hanawa  2005. This  claimed, using
a grid based code, that actually  material of high specific
angular momentum is preferentially accreted on to the binary {\it primary}
and thus that the mass ratio should {\it fall}. This study
did not contradict the widely  acknowledged result that
the flow preferentially enters the secondary's Roche lobe, but argued instead 
that, after half an orbit around the secondary, it crosses via the L1 point
into the primary's Roche lobe and is then accreted. Ochi et al conjectured that
the failure of SPH calculations to demonstrate this behaviour stemmed from
the excessively viscous nature of SPH, which might cause  the SPH particles to
spiral in sufficiently, in half an orbit, so that they could avoid the L1 region.
In the same spirit, proponents of  SPH  countered that Ochi et al's
flows were too warm (with sound speed of $0.25 \times$ the orbital velocity of the binary) and that this caused artificial
acceleration of the flow through the L1 point. Clearly these issues had to be
investigated further, using both Lagrangian and Eulerian codes. At the time of 
his death, Eduardo was close to completing a suite of high resolution SPH
calculations which were beginning to illuminate the problem and whose results
I summarise below.

 His  repetition, at higher resolution,  of the SPH calculation of Bonnell \& Bate 1997 
(which modeled  accretion onto
a protobinary of a flow with fixed specific angular momentum )
demonstrated that the rate of increase of q 
is indeed somewhat resolution dependent and is over-estimated
at low resolution (about  $5 \times 10^4$  particles in the discs) compared with the highest
resolution (a hundred times more particles, which has achieved numerical convergence)
by about a factor two. This can be traced to the fact that at low resolution,
particles in the outer part of the primary's Roche lobe are not behaving
in a fuly fluid-like manner and, through precession of their elliptical orbits,
are able to tip back through the L1 point into the secondary's Roche lobe.
Nevertheless, q  {\it increases} (for this value of specific angular
momentum of the flow) {\it at all resolutions},
 in contrast to the results of Ochi
et al.

 Eduardo found that at high resolution, the shocks are well  delineated 
(see left hand panel of Figure 2) and, indeed, 
that the Jacobi constant (i.e. the Bernoulli function in the co-rotating
frame) is well  conserved along streamlines except in shocks. 
This result goes against the hypothesis
of Ochi et al that the SPH results are driven by artificial viscosity in
the secondary's disc flow, since under these cirumstances the action of artifical viscosity should cause changes in the Jacobi constant, even away from shocks.
Crucially, 
however,  the shock geometry is quite different in Eduardo's simulations
compared with those of Ochi et al.  In the
SPH calculation, loosely bound material in the secondary's Roche lobe
encounters a shock with material flowing in from the direciton of the primary
(see dense structure to the East of the secondary in left hand panel of
Figure 2) and it is dissipation in this shock which prevents material flowing through
L1 and into the primary's Roche lobe. This shock is simply not there in Ochi et al's caclulation, so that material entering via L2 retains enough energy to 
smoothly transit through L1.

\begin{figure}[!ht]
\plottwo{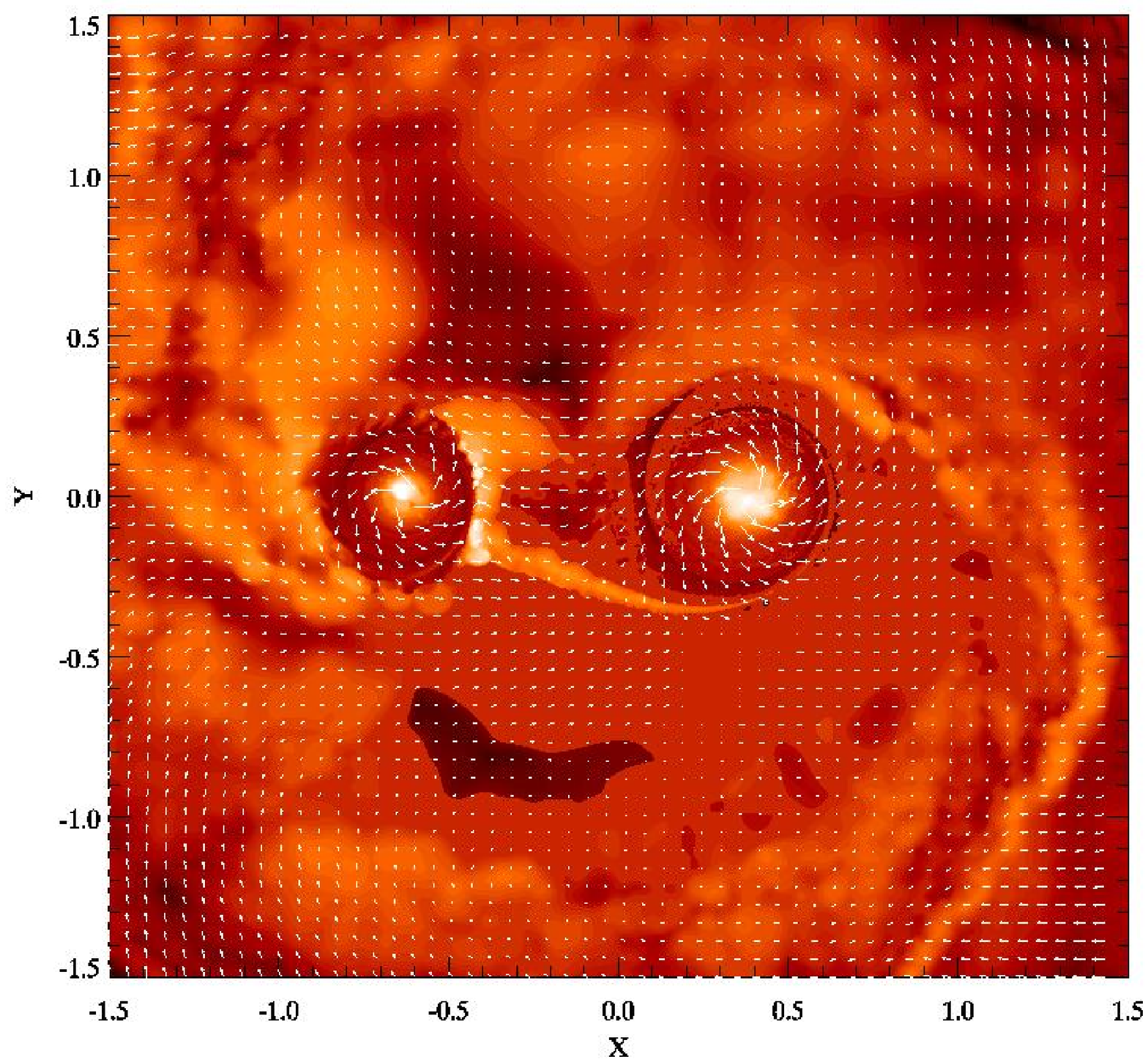}{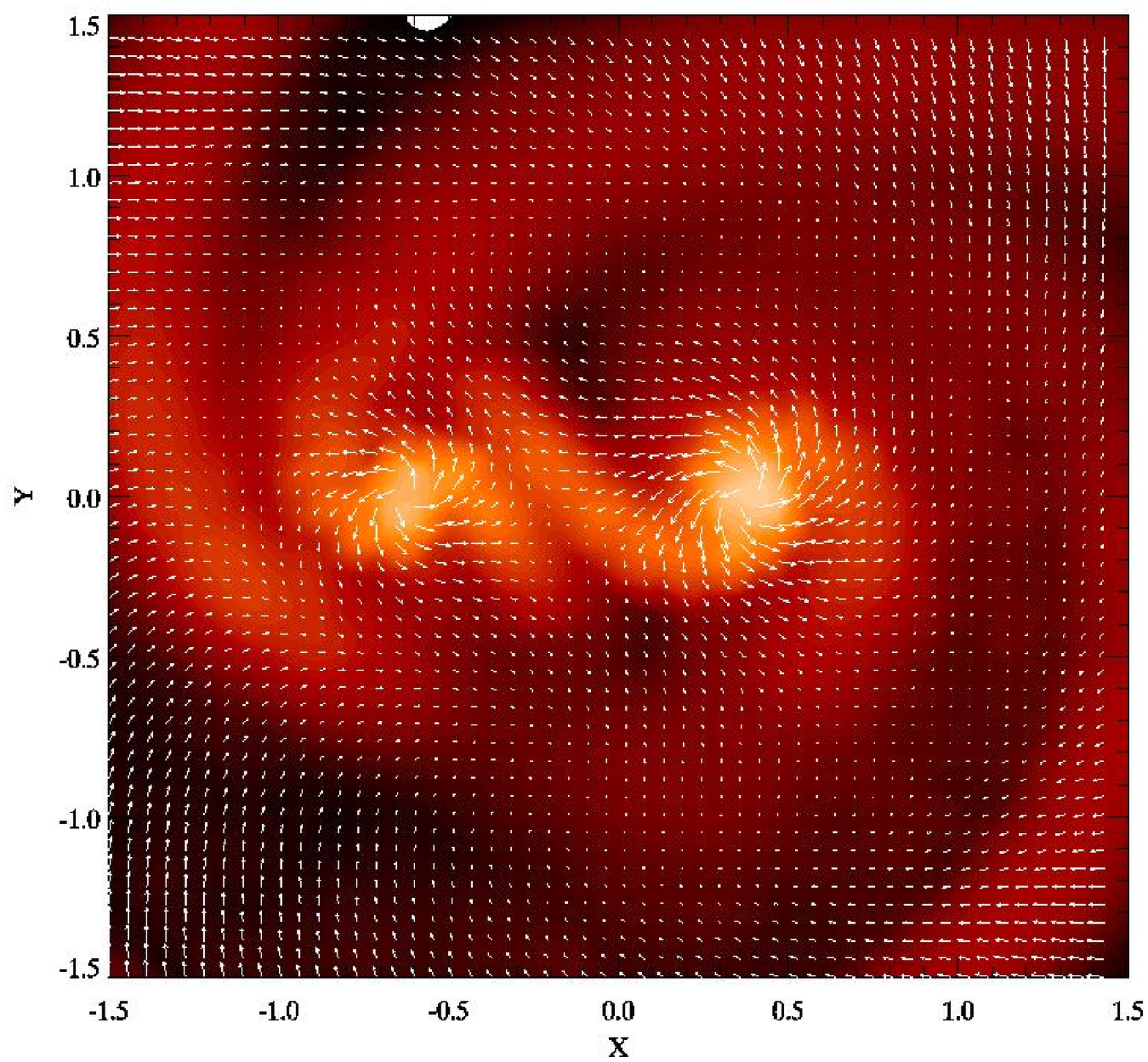}
\caption{Density and flow structure in the co-rotating frame
around an accreting proto-binary,
contrasting  a cold three dimensional simulation (left) and
a warm two dimensional simulation (right). The secondary is to the left.
Note the pronounced shock to the East  of the secondary in the left hand panel.
Unpublished
simulations by E. Delgado}.
\end{figure}
  Thus  while both simulations are apparently
self-consistent, {\it given the location of the shocks},   one has to 
enquire why the shock morphology is so different in the two cases. One possibility
is that - although the SPH and Eulerian simulations are matched in terms of
specific angular momentum of the accretion flow and mass ratio of the
binary - there are two potentially important differences. Firstly, 
the Eulerian calculation is much warmer (about a factor $25$ in
temperature). Secondly, it  is strictly two dimensional
whereas the SPH calculation  - although highly flattened near 
the binary - is fully three dimensional and introduces particles on the surface
of a sphere at large radius. An indication that one, or both, of these
effects may be important is provided by one of Eduardo's simulations which
mimicked as closely as possible the precise conditions of Ochi et al, being
both 2D and warm. The 
right hand panel of Figure 2  demonstrates that the shock has in this case moved away from
the secondary's Roche lobe, in qualitative agreement with the flow morphology
of Ochi et al. Although {\it quantitative} agreement between the
SPH and Eulerian codes is  not achieved even here, the {\it sign}
of the change is encouraging. This shift in flow morphology is 
reflected in a change in the relative accretion rates on to the primary
and secondary. Although  q  still increases in these
warm 2D SPH calculations, 
$\dot q$ is reduced by an order of magnitude compared with the cold 3D
calculation.    

   This was the situation at the time of Eduardo's death and there is
evidently still much for his collaborators   to do.
We now suspect 
that  gas temperature and/or flow dimensionality  are important
determinants of the shock  morphology, and we must now 
discover which of these two effects is the critical one. One then has to
judge  which set of conditions is  more realistic.  
Ultimately, this will tell us whether the old `simple'
idea is correct - i.e. that the flow simply enters the secondary's Roche
lobe and remains therein - or whether the more complex outcome of Ochi
et al - involving flow between Roche lobes - is closer to the truth.

  All this has implications for the extreme mass ratio binary problem which 
Eduardo had been
trying to solve for years.
If the result is closer to that of Ochi et al, then this
could solve the problem at a stroke. If the results instead support
the old SPH results, then the puzzle remains (although 
if low resolution simulations have over-estimated the
{\it magnitude} of the growth of q, then  
binaries may remain at low q, even if the sign of $\dot q$ is positive).  
Otherwise, one has to think harder about how to  prevent
the infall of gas with high specific angular momentum onto
protobinaries. Although it is tempting to invoke feedback for   this, 
it is not obvious why this should
be so important in the low mass binaries which we are largely trying
to explain.

\section{Eduardo the person}

Having talked  about the problems that motivated Eduardo as
a scientist, it
leadss me  to how much can one discern of Eduardo the person in all
this? An unexpected death  leads one to analyse what one saw of
a person in the daily contacts of studentship and collaboration. Eduardo was
always calm, always gentle, always attentive. He was never arrogant, never
distorted the problem at hand, or his relations with those around him, by
the obtrusion of his own ego.  He commanded the respect of
all those who came into contact with him through this obvious intellectual
and personal integrity.  He was a joy to work with.

  I see great continuity between those characteristics as a collaborator
and the personal testimonies  from his friends, many of which
are movingly collected on his blogspot (http://eduardo-delgado.blogspot.com/)
I was aware that Eduardo was
very popular with his peers in Cambridge and that he belonged to a big
crowd of friends of all nationalities whose joyful social life
never got in the way of Eduardo's scientific activities. But these messages
- and the evidence of the conversations I have had with so many - speak
of a very special warmth of affection and regard. Unfortunately I can only
dimly decipher the majority in Spanish but they clearly tell the same story
as those I can read (in Italian and English). So many speak of the gentleness
of his eyes, the warmth of his friendship, his calmness (one Italian friend
exclaimed `You even played football  with calm' ( incredible!)). He
was, to put it simply, much loved.

  All this is very particular to Eduardo, but I should  end on a note that
perhaps has relevance to all those gathered for this meeting and to
astronomers everywhere. In the last months of his life,
Eduardo was particularly happy. A former Cambridge colleague,
who stayed with him on an observing run at Christmas time, told me `You
know, Edu was just {\it so} happy when I last saw him'. He had everything he
dreamed of, living in the island he loved, starting new projects in his
new postdoc with Casiana Munoz, also continuing his independent research
(he felt he was so  close to cracking the extreme mass ratio binary problem).
We're perhaps not all in one of those special moments in life when everything
seems to be working, but I reckon that if any one of us suffered the same
awful and random misfortune as Eduardo, then they would be able to say of
us that we'd spent our lives doing what we loved. This is  a rare
privilege. So, in keeping with the celebratory mood of this conference, let us
recall  
that, while mourning Eduardo no less, we can also
celebrate a life spent - like ours - in doing something we believe to
be intrinsically valuable. To do this in a spirit of calm and integrity
was Eduardo's gift. We miss him immensely.

\begin{figure}[!ht]
\plotone{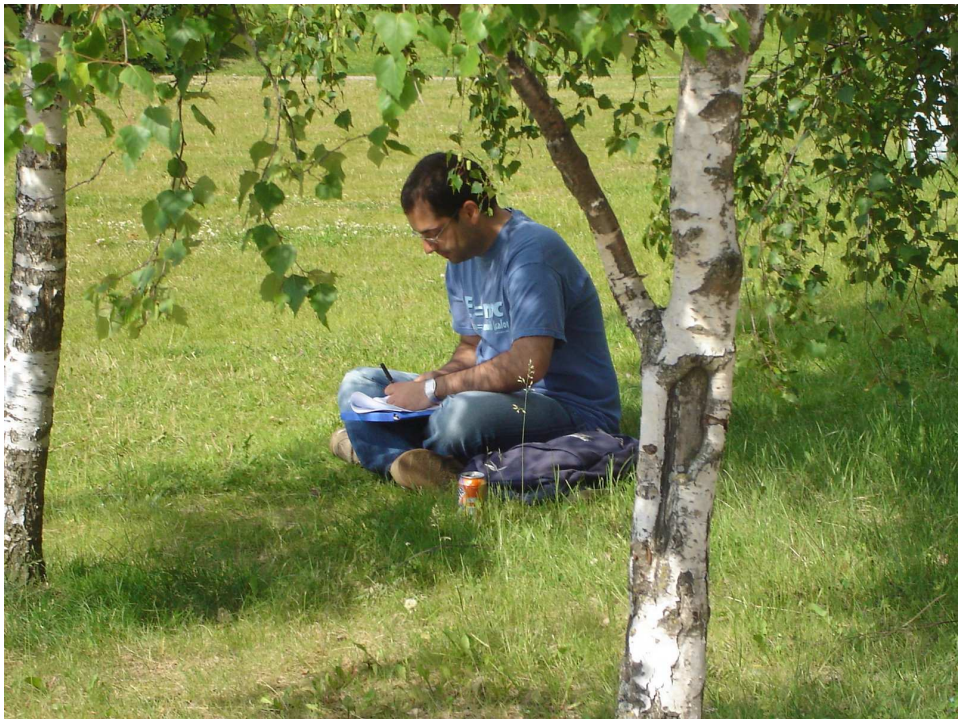}
\caption{Edu emanating his characteristic calm}
\end{figure}

{}


\begin{thebibliography}{}
\bibitem[Artymowicz (1983)]{}
Artymowicz, P. 1983 Acta Astron. 33,223\\
\bibitem[Delgado et al (2004)]{}
Delgado-Donate, E. J., Clarke, C. J., Bate, M. R. \& Hodgkin, S. T. 2004, MNRAS 351,617\\
\bibitem[Duquennoy \& Mayor (1991)]{}
Duquennoy, A. \& Mayor, M. 1991, A \& A 248,485\\
\bibitem[Bate (2000)] {}
Bate, M. R. 2000, MNRAS 314,33\\

\bibitem[Bate \& Bonnell (1997)]{}
Bate, M. R \& Bonnell, I. A. 1997, MNRAS 285,33\\
\bibitem[Bouy et al (2003)]{}
Bouy, H., Brandner, W., Martin, E., Delfosse, X., Allard, F. \& Basri, G. 2003, AJ 126,1526
\bibitem[Mazeh et al (1992)]{}
Mazeh T., Goldberg D., Duquennoy A., Mayor M. 1992, ApJ, 401, 265\\
\bibitem[Garcia \& Mermilliod]{}
Garcia, B. \& Mermilliod, J. 2001,  A \& A 368,122\\
\bibitem[Ochi et al (2005)]{}
Ochi, Y., Sugimoto, K. \& Hanawa, T. 2005, ApJ 623,922\\
\bibitem[Tokovinin et al (2006)]{}
Tokovinin, A., Thomas, S., Sterzik, M. \&  Udry, S. 2006, A \& A 450,681
\end{thebibliography}
\end{document}